\documentclass[12pt]{article}

\usepackage{amssymb}
\usepackage{authblk}
\usepackage{comment}
\usepackage{amsmath,empheq,systeme}
\usepackage{mismath}
\usepackage{xcolor}
\usepackage{graphicx}
\usepackage{caption}
\usepackage{subcaption}
\usepackage{siunitx}
\DeclareSIUnit\bar{bar}
\usepackage[version=4]{mhchem}
\usepackage{booktabs}
\usepackage{multirow}
\usepackage[numbers,sort&compress]{natbib}
\newcommand{\keps}{$k$-$\varepsilon$ }

\usepackage[english]{babel}
\usepackage{latexsym}

\usepackage{url}
\usepackage{xcolor}
\definecolor{newcolor}{rgb}{.8,.349,.1}
\usepackage{hyperref}

\title{Validation of a Reynolds-averaged numerical simulation environment to simulate high-pressure, auto-igniting hydrogen diffusion flames}%

\author{%
N. Diepstraten\thanks{Corresponding author. E-mail: \url{n.diepstraten@tue.nl}},\,
L.M.T. Somers,\, 
J.A. van Oijen
}

\affil{\textit{Department of Mechanical Engineering (Power \& Flow Group),\\
Eindhoven University of Technology,\\
Groene Loper 3, 5612 AE Eindhoven,\\
Noord-Brabant, The Netherlands}}

\date{} 

\begin{document}
\maketitle
\vspace{-1.5em}

\begin{abstract}
The hydrogen (\ce{H2}) fueled direct-injection (DI) compression-ignition (CI) argon power cycle (APC) is an attractive technology to counteract the mismatch between energy demand and supply from renewable energy sources. 
The development of the APC, as well as air-breathing DI CI \ce{H2} engines, can be advanced by computational fluid dynamics (CFD) models. 
Reynolds-averaged Navier-Stokes (RANS) combustion models are an effective approach to predict global in-cylinder variables such pressure and heat release rate. 
However, validity of this method to simulate the complex phenomena associated with high-pressure, auto-igniting hydrogen jets is not ensured due to the underlying model assumptions.
In this study, a RANS CFD environment using two commonly used eddy-viscosity models and the Reynolds stress model (RSM) is extensively validated. 
Combustion is modeled with a detailed chemistry model.
First, hydrogen distributions in non-reacting jets are compared against literature data to assess the accuracy of the models on turbulent mixing at high pressure. 
Subsequently, the capability of the model to simulate auto-igniting jets is assessed by comparison with reported measurement data of closed-vessel experiments at high pressure.
Ignition delay times, pressure rise profiles, and heat release rate profiles are compared for different ambient temperature and oxygen concentration. 
Adequate agreement is found for the diffusive combustion phase for all models, despite the lack of turbulence-chemistry interaction in the combustion model.
Trends with ambient temperature and oxygen concentration were well predicted and the best agreement is found for the RSM.
\end{abstract}

\section*{Novelty and significance statement}
In this study, for the first time a widely used RANS CFD environment is extensively validated for simulation of high-pressure, auto-igniting hydrogen jets encountered in direct-injection compression-ignition hydrogen engines.
Simulation results obtained with three commonly used turbulence models are compared with experimentally measured 2D mole fraction fields, as well as pressure rise curves and derived heat release rate profiles at various conditions. 
We identify strengths and weaknesses of the CFD environment to simulate complex combustion phenomena, such as the transition from premixed to diffusive combustion mode. 
Thanks to the comparison with different experimental measurements, this study increases the understanding and provides valuable insights into the capabilities and limitations of the RANS CFD environment to predict important characteristics such as jet penetration and heat release rate, and thereby aids the development of efficient hydrogen engines.

\section{Introduction}
\label{sec:introduction}
The hydrogen (\ce{H2}) fueled internal combustion engine (ICE) is an attractive technology to counteract the intermittent energy supply by renewable sources, thanks to its quick response time and high power density.
Disadvantages of the ICE compared to fuel cells are its nitrogen oxide emissions and lower efficiency.
The latter can be alleviated by operating at high compression ratios, which can only be realized with direct-injection (DI) compression-ignition (CI) engines.
DICI \ce{H2} engines are challenging, because high in-cylinder temperatures are needed to achieve sufficiently short ignition delay times due to hydrogen's high auto-ignition temperature compared to diesel~\cite{yip2019review}.

One option to tackle this challenge is to increase the compression ratio of the engine, to assist ignition with a glow or spark plug~\cite{homan1979,homan1983}, or with a pilot combustion of a more reactive fuel such as diesel~\cite{trusca2001high,mctaggart2015direct}. 
Another option is to change the atmospheric nitrogen for a mono-atomic gas, such as argon, resulting in a combustion concept to which we will refer as the argon power cycle (APC).
An advantage of argon compared to nitrogen as working fluid is that it has a lower heat capacity thanks to the absence of rotational and vibrational degrees of freedom. 
The lower heat capacity of argon results in higher in-cylinder temperatures and pressures.
This makes the APC suitable for a DICI \ce{H2} combustion strategy. 
Another main advantage of argon's low heat capacity is that the theoretical (Otto-cycle) efficiency increases considerably.
This has been demonstrated in experimental studies as well~\cite{kuroki2010study,killingsworth2011increased,aznar2017experimental,ikegami1982study}.
Moreover, there is no nitrogen oxide formation due to the absence of nitrogen in the reactants, which makes the APC an emission free energy conversion system in theory. 
The only constituents of the exhaust stream are argon and water in case of stoichiometric combustion.
If the water in the exhaust is recovered by condensation, the argon can be reused through a closed-loop configuration \cite{kuroki2010study,killingsworth2011increased,aznar2017experimental,sierra2018experimental,ikegami1982study,wang2023future}.

To develop such engines, accurate and computationally efficient computational fluid dynamics (CFD) models are needed to reduce expensive prototyping and to gain insights in in-cylinder processes which cannot be obtained easily via experiments.
Adequate simulation of the injection and combustion process is crucial to predict global in-cylinder variables, such as pressure, temperature and heat release rate and to determine engine performance indicators, such as thermal efficiency and heat loss.
The in-cylinder flow can reach Reynolds numbers $\mathit{Re} > 10^6$ during the injection. 
Modeling of these flows requires the use of a coarse scale method, such as Reynolds-averaged Navier-Stokes (RANS) or large-eddy simulation (LES), since resolving all scales via direct numerical simulation is computationally unfeasible.

The RANS method has been the mainstay for engine research and development for decades thanks to its ability to predict trends of mean flow quantities and computational efficiency.
However, the continuously increasing computing power has enabled LES's of engine-cycles as well~\cite{rutland2011large}.
Since an LES solves for most of the turbulent structures, it can accommodate more detailed analyses, e.g.\@ regarding mixing processes and cycle-to-cycle variations.
On the other hand, the computational costs associated with an LES are considerably larger than for a RANS simulation.
For this reason, it is believed that RANS will remain important for engine research~\cite{posch2025turbulent,yang2014rans}. 
It is therefore relevant to validate a RANS CFD environment that can adequately predict the injection and subsequent combustion process in DICI \ce{H2} engines.

The combustion process of DICI \ce{H2} engines is preferred to be predominantly non-premixed, since a large amount of premixed combustion would result in a high pressure rise rate which can damage the engine.
For turbulent non-premixed combustion, the rate of combustion is mainly determined by turbulent mixing as the chemical time scales are much smaller~\cite{peters2000turbulent}, which is why it is sometimes referred to as mixing-controlled combustion.
Thus it is important that the RANS CFD environment can accurately predict the spatial development of the hydrogen jet and the mixing process with its surrounding oxidizer.
These RANS CFD environments typically use approximations to model complex phenomena, such as turbulence and combustion.
The applicability of these closure models to hydrogen is not ensured, since hydrogen's physical properties largely deviate from other commonly used fuels.

A pioneering study of numerical modeling of hydrogen jets for engine applications was already performed in 1995 by Johnson et al.~\cite{johnson1995three}.
They used the standard \keps model and a near-nozzle grid resolution of 2 grid cells per nozzle diameter and found that the jet penetration was severely under-estimated.
Refining the grid near the nozzle exit helped, but this led to impractically high computational costs at that time.
The agreement with experiments was improved after modifying the turbulence model.
Furthermore, the experimental auto-ignition delay time could not be reproduced.
Also Ouellette~\cite{ouellette1996direct} observed an under-penetration of under-expanded methane jets using the same turbulence model, despite the use of a higher grid resolution (up to 4 grid cells per nozzle diameter).
As a consequence, the spreading rate was over-estimated. 
The differences with the experimental data were attributed to the isotropic nature of the \keps model. 
In~\cite{keysar2004numerical}, different variants of the isotropic \keps model and a non-isotropic Reynolds stress model (RSM) simulating correctly, under-, and over-expanded air-in-air jets were compared against velocity experiments. 
Accurate results were obtained with the RSM, but the solution would not converge for highly under-expanded jets. 
The standard \keps and Renormalization Group (RNG) models provided satisfactory agreement as well.
Interestingly, the authors reported that there was no need for a round-jet correction~\cite{keysar2004numerical}.
In a similar study~\cite{miltner2015cfd}, it was concluded that the RSM resulted in the highest accuracy, but also the standard \keps model served well. 
Babayev et al.~\cite{babayev2021computational} simulated non-reacting jets with various \keps and $k$-$\omega$ models, and compared the results with oil mist scattering experiments at three different pressure ratios. 
They reported that the \keps RNG model gave the most consistent and realistic results based on jet penetration and shape, but the results of the different turbulence models were not presented. 
Besides, insights in the spatial distribution of the hydrogen jet were not obtained, and no validation was done for igniting hydrogen jets. 
In summary, it remains unclear how to setup a RANS CFD environment that is valid for high-pressure, auto-igniting hydrogen jets encountered in DICI \ce{H2} engines.

In the present study, we address this literature gap by setting up a model using a widely used CFD environment and validating it against experiments.
Based on the findings in the literature reviewed above, three turbulence models are selected: the standard \keps model, the \keps RNG model and the RSM.
Details of the CFD setup are provided in Section~\ref{sec:simulation_setup}.

In Section~\ref{sec:non_reacting_jets}, the simulation results are compared to inverse laser induced fluorescence (ILIF) measurements of inert hydrogen jets in a pressurized nitrogen environment reported in~\cite{peters2025thesis,peters2025lif}.
Since flame-wall interaction largely influences heat loss and hence the thermal efficiency of a DICI \ce{H2} engine \cite{babayev2021computational,yip2019review}, it is important to accurately simulate the penetration and shape of the hydrogen jet.
For this reason, the simulation results will be assessed based on distributions of hydrogen mole fraction and jet penetration curves.
Furthermore, the accuracy of the selected turbulence models in predicting the mixing behavior will be investigated by comparing radial profiles of hydrogen mole fraction with experimental results. 
In Section~\ref{sec:reacting_jets}, the numerical setup's capability to simulate auto-igniting hydrogen jets encountered in air-breathing DICI \ce{H2} engines and APC is assessed.
The pressure rise due to combustion directly translates to the work done by the engine.
For this reason, pressure rise traces resulting from the constant volume chamber experiments reported in literature~\cite{peters2025thesis,peters2025ehpc} are compared to the numerical counterparts.
To gain more insights in the combustion process, heat release profiles are processed from the pressure traces.
Besides variation of the background gas to ensure the model's validity for air-breathing DICI \ce{H2} engines as well as the APC, variations in ambient temperature and oxygen concentration are investigated.
The latter is particularly interesting for the APC, as the oxygen concentration is an additional design parameter compared to air-breathing engines.
Lastly, conclusions will be drawn in Section~\ref{sec:conclusions}.

\section{Simulation setup}
\label{sec:simulation_setup}

\subsection{Physical modeling}
\label{subsec:governing_equations}
The commercial software package CONVERGE version 3.0 \cite{converge} is used to solve the compressible RANS equations.
In this study, two first-moment closures to model the Reynolds stresses are investigated: the \keps model of Sarkar et al.~\cite{sarkar1991analysis}, to which we will refer as the standard \keps model, and the \keps RNG model of Yakhot et al.~\cite{yakhot1992development}.
It is good to note that the \keps RNG model, first described in \cite{yakhot1986renormalization}, contains a low-Reynolds number modification, but only the high-Reynolds number limit is implemented in CONVERGE (see \cite{yakhot1992development,converge_manual}).
It includes an additional term in the equation for the turbulent dissipation ($\varepsilon$) that is neglected in the standard \keps model to improve model predictions of highly strained flows.
Besides, the model constants were derived by the renormalization group theory and are therefore not experimentally adjustable.
In addition to these eddy-viscosity models, the Launder-Reece-Rodi (LRR) RSM is investigated~\cite{gibson1978ground}.
This model allows for anisotropic turbulence, which may result in more accurate results especially in the vicinity of the nozzle due to the high local shear gradients.

Combustion is modeled with the SAGE detailed chemistry solver~\cite{senecal2003multi}.
The chemistry is directly integrated using the reaction mechanism of Burke et al.~\cite{burke2012comprehensive}, which is validated for high-pressure \ce{H2}/\ce{O2} flames.
No turbulence-chemistry interaction (TCI) model is included.
From a theoretical perspective, turbulent fluctuations should not be ignored due to the non-linearity of the chemical source terms resulting in a commutation error.
However, from a practical point of view, there is no consensus yet whether TCI is essential for accurate engine simulations exhibiting mixing-controlled combustion~\cite{posch2025turbulent}.
An argument found in literature to justify the neglection of TCI is that small-scale interactions between turbulence and chemistry may not play a dominant role in the overall combustion rate \cite{kokjohn2011investigation,posch2025turbulent}.

To model hydrogen's real gas behavior, the Soave modification of the Redlich-Kwong equation of state is used~\cite{soave1972equilibrium}. 
As reported in~\cite{babayev2021computational}, it is important to include species dependent critical properties. 
The values are obtained using thermodynamic library CoolProp~\cite{coolprop}.
The thermodynamic fluid properties are modeled using NASA polynomials.
A mixture-averaged diffusion model is used to calculate the molecular diffusion constants.
The turbulent Prandtl and Schmidt numbers are both set to 0.7.

\subsection{Computational domain and boundary conditions}
\label{sec:computational_domain}
\begin{figure}
  \centering
  \includegraphics[width=0.8\textwidth]{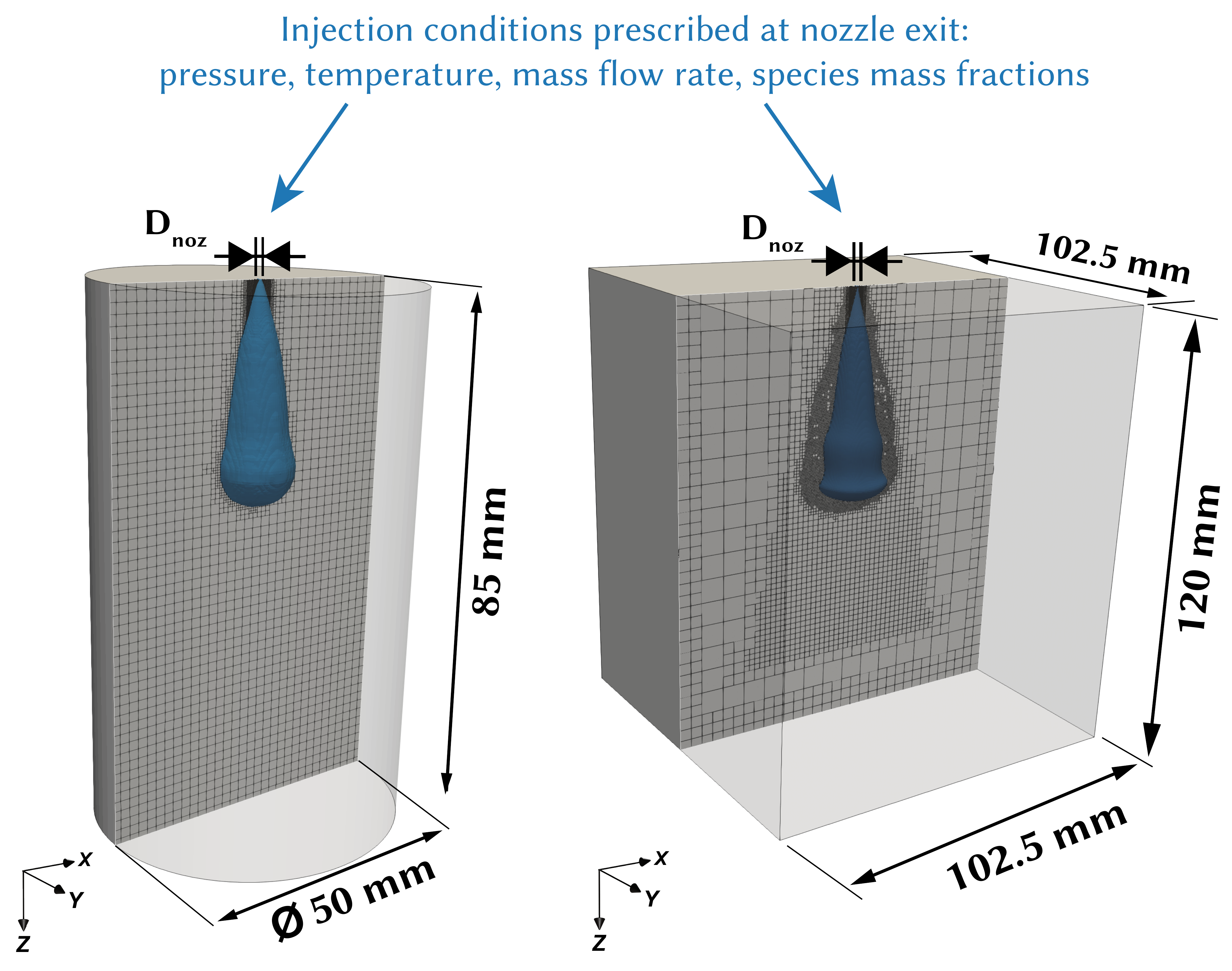}
  \caption{Computational domain used for the non-reacting jets on the left and for the reacting jets on the right. The blue surfaces project the jet shape.}
  \label{fig:computational_domains}
\end{figure}

The computational domain for the non-reacting hydrogen jets is defined as a cylindrical vessel with a circular inflow boundary located at the center of the top wall representing the nozzle exit plane. 
The imposed conditions at this plane will be detailed in the next section.
Figure~\ref{fig:computational_domains} shows a render of the domain and its main dimensions.
The top of the vessel is modelled using the law of the wall, and its temperature is equal to the initial temperature of the gas inside the vessel.
The cylindrical and bottom boundaries of the vessel are modelled with an outflow boundary condition to maximize the free propagation of the jet.

\begin{figure}
  \centering
  \includegraphics[width=0.6\textwidth]{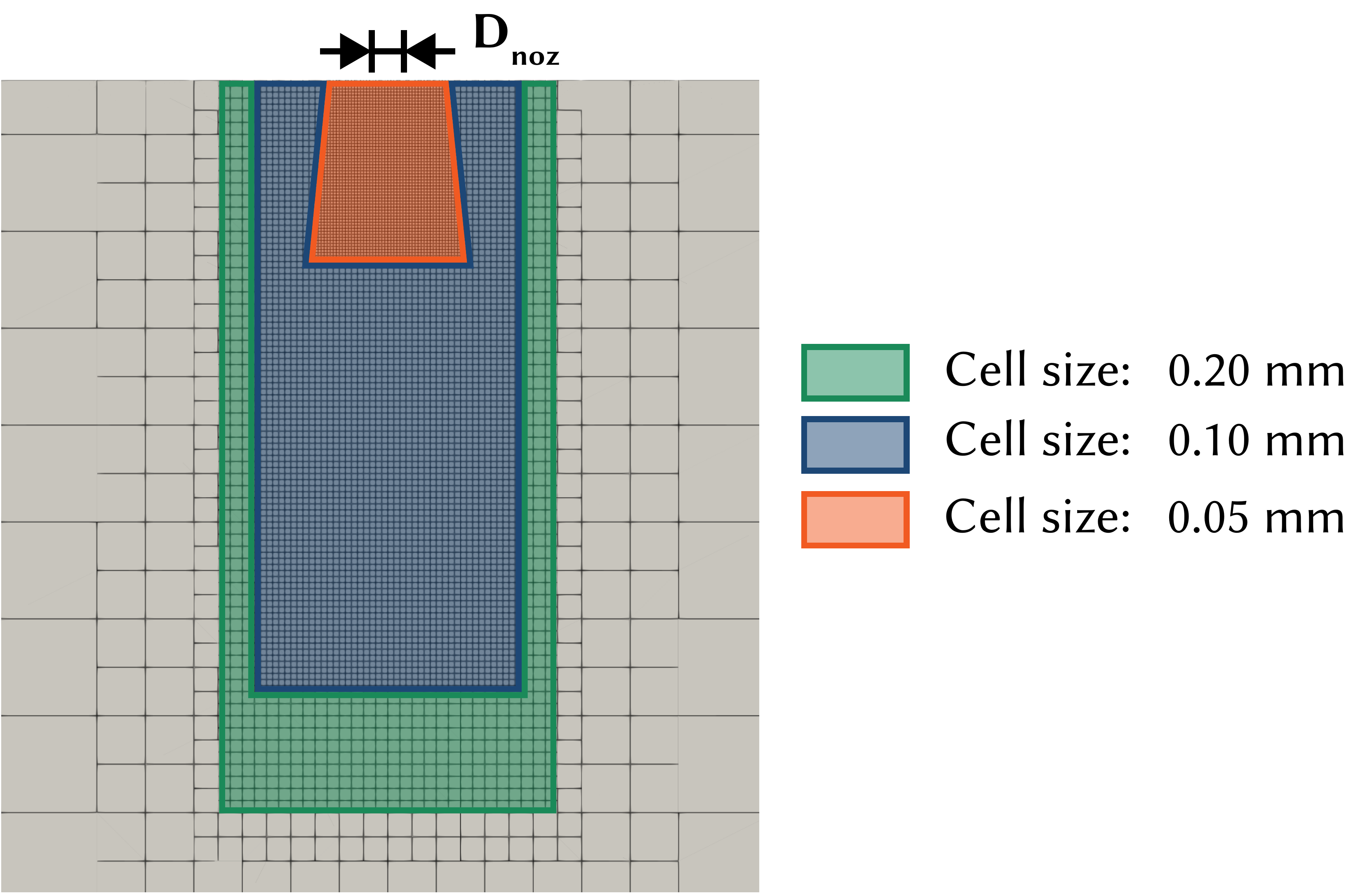}
  \caption{Zoom of the computational domains showing local grid refinements in the vicinity of the nozzle.}
  \label{fig:grid_refinements}
\end{figure}

The experiments involving reacting jets are performed in a closed vessel.
To directly compare the pressure trace of the experiment with the simulation, a second computational domain is defined by a vessel having the same internal volume as its experimental counterpart and approximately the same dimensions. 
The injection is again realized by a circular inflow boundary which represents the nozzle exit plane. 
A circular boundary around the nozzle exit plane is modelled using the law of the wall and a temperature of \SI{383}{\K}.
The inclusion of this relatively cold wall around the injection boundary in the numerical setup mimics the injector tip which is kept at \SI{383}{\K} in the experiments.
The rest of the boundaries are modelled as adiabatic walls.
The domain, including its dimensions, is shown in Fig.~\ref{fig:computational_domains}.

The base grid size is set to \SI{1.6}{\milli\m} in both setups.
To expedite the computations for the closed vessel simulations, the grid is coarsened to \SI{6.4}{\milli\m} outside the projected jet area, as can be seen in Fig.~\ref{fig:computational_domains}.
In both domains, local grid refinements are added in the vicinity of the nozzle to adequately handle the large gradients, as indicated in Fig.~\ref{fig:grid_refinements}.
Additionally, adaptive mesh refinement (AMR) is used to locally increase the grid resolution to \SI{0.4}{\milli\m} to ensure that large gradients in velocity and temperature are resolved adequately according to the method of~\cite{bedford1993conjunctive,pomraning2000development}.
Besides, regions in which the hydrogen mass fraction ranges between \SI{1e-3}{} and \SI{1e-4}{} are also refined to \SI{0.4}{\milli\m} to capture hydrogen's high diffusivity as recommended by Babayev et al.~\cite{babayev2021computational}.
This grid strategy is similar to the one described in \cite{babayev2021computational}, in which an extensive grid study was performed for high-pressure hydrogen injections.

\subsection{Numerics}
The RANS equations are solved by the density-based Pressure-Implicit Splitting of Operators (PISO) solver of Issa et al.~\cite{issa1991solution}. 
The convective fluxes are discretized using a blending scheme, which uses a second-order central-differencing or first-order upwind scheme depending on the local monotonicity of the flow to ensure stability. 
For the equations for the turbulent kinetic energy $k$ and its dissipation $\varepsilon$, a first-order upwind scheme is used.
A first-order Euler approach is used for time-integration. 
The time-step is adapted during the simulation and is limited by a Courant-Friedrich-Lewis (CFL) number of 0.6 for velocity and by a maximum time-step of \SI{1e-6}{\second} after the end of injection.

\subsection{Nozzle conditions}
\label{sec:injection_conditions}
The injection is realized by defining an inflow boundary condition at the nozzle exit plane by imposing a mass flow rate, pressure and temperature as function of time. 
The corresponding values are computed using the global conservation model (GCM) introduced in our previous work~\cite{diepstraten2025gcm}.
A constant total temperature of the hydrogen is assumed in the injector, which was shown to give accurate results.
All jets investigated in this study have an injection pressure of \SI{100}{\bar}.
It is good to note that the nozzle pressure ratio (NPR) of the simulated conditions is at least 2.5.
Since the critical NPR of hydrogen for choked flow is NPR $\approx1.9$ \cite{shapiro1953dynamics}, the nozzle exit conditions are independent of the conditions downstream.
However, a different injector is used for the non-reacting jet experiments than for the auto-igniting jet experiments, leading to different nozzle diameters and discharge and momentum coefficients.
The injector properties and nozzle exit conditions are summarized in Table~\ref{tab:nozzle_conditions}, where $d_\mathrm{noz}$ is the nozzle diameter.
It was measured in experiments that it takes $\sim$\SI{0.1}{\milli\s} to retract the needle of both injectors~\cite{peters2025thesis,peters2026characterization}.
This transient process is mimicked in the injection profiles by a linear transition from chamber conditions at $t=0$ to injection conditions at $t=\SI{0.1}{\milli\s}$.
\begin{table}
	\centering
	\caption{Injector properties and nozzle exit conditions}
    \label{tab:nozzle_conditions}
	\begin{tabular}{lll}
		\hline
		 & Non-reacting jets & Reacting jets \\ \hline
		$d_\mathrm{noz}$ [\unit{mm}]      & 0.65  & 0.55 \\
		$C_\mathrm{D}$ [-]                & 0.44  & 0.58 \\
		$C_\mathrm{M}$ [-]                & 0.74  & 0.91 \\
        $\dot{m}$ [\unit{\g\per\s}]       & 0.924 & 0.763 \\
		$p$ [\unit{\bar}]                 & 8.57  & 12.8 \\ 
    $T$ [\unit{\K}]                   & 149   & 205 \\ \hline
	\end{tabular}
\end{table}

\subsection{Vessel conditions}
The non-reacting jets were injected in an environment composed of nitrogen and acetone as LIF tracer gas at \SI{293}{\K} and \SI{10}{\bar}, resulting in an NPR of 10. 
Since the mass fraction of tracer gas in the vessel is small and acetone's molecular weight is comparable to that of nitrogen, no acetone is introduced in the simulation domain.

The experiments concerning the auto-igniting hydrogen jets were all conducted at a chamber pressure of \SI{40}{\bar}, resulting in an NPR of 2.5. 
In this study, we assess the performance of the CFD model to simulate auto-igniting hydrogen jets in \ce{Ar}/\ce{O2} and \ce{N2}/\ce{O2} environments by varying the temperature and oxygen concentration in the vessel, thereby investigating six conditions in total.
The conditions in the vessel are summarized in Table~\ref{tab:ehpc_vessel_conditions}.
Since the effect of turbulent kinetic energy in the vessel and the nozzle on jet development is negligible~\cite{ouellette1996direct}, $k$ and $\varepsilon$ are initialized with values close to zero for all simulations.
\begin{table*}
  \centering
  \caption{Vessel conditions before the hydrogen injection for the reacting jets}
  \label{tab:ehpc_vessel_conditions}
  \begin{tabular}{lcccccc}
    \hline
    \multirow{2}{*}{Label} & \multicolumn{2}{c}{Baseline} & \multicolumn{2}{c}{Higher $T$} & \multicolumn{2}{c}{Lower $X_{\ce{O2}}$} \\ 
     & \ce{N2} & \ce{Ar} & \ce{N2} & \ce{Ar} & \ce{N2} & \ce{Ar} \\ \hline
    $p$ [\unit{\bar}] & \multicolumn{2}{c}{40} & \multicolumn{2}{c}{40} & \multicolumn{2}{c}{40} \\
    $T$ [\unit{\K}] & \multicolumn{2}{c}{1200} & \multicolumn{2}{c}{1400} & \multicolumn{2}{c}{1200} \\
    $X_{\ce{O2}}$ [\unit{\percent}] & \multicolumn{2}{c}{15} & \multicolumn{2}{c}{15} & \multicolumn{2}{c}{10} \\
    $X_{\ce{N2}}$ [\unit{\percent}] & 71.5 & 0.0 & 71.4 & 0.0 & 76.1 & 0.0 \\
    $X_{\ce{Ar}}$ [\unit{\percent}] & 3.6 & 77.8 & 3.7 & 77.8 & 4.1 & 83.2 \\
    $X_{\ce{CO2}}$ [\unit{\percent}] & 6.6 & 4.8 & 6.6 & 4.8 & 6.6 & 4.6 \\
    $X_{\ce{H2O}}$ [\unit{\percent}] & 3.3 & 2.4 & 3.3 & 2.4 & 3.3 & 2.3 \\ \hline
    \end{tabular}
\end{table*}

\section{Non-reacting jets}
\label{sec:non_reacting_jets}
In this section, we discuss the performance of the selected RANS turbulence models based on non-reacting hydrogen jets. 
For this purpose, the simulation results obtained with the computational domain in Fig.~\ref{fig:computational_domains} are compared to inverse ILIF measurements described in~\cite{peters2025thesis,peters2025lif}.
Note that the experimental images are ensemble averaged, while the RANS equations are Favre-averaged.
However, the difference between the ensemble averaged and Favre averaged experimental images are found to be very small ($<$\SI{1}{\percent}).

\begin{figure*}
    \begin{subfigure}[t]{0.32\textwidth}
      \includegraphics[width=\textwidth]{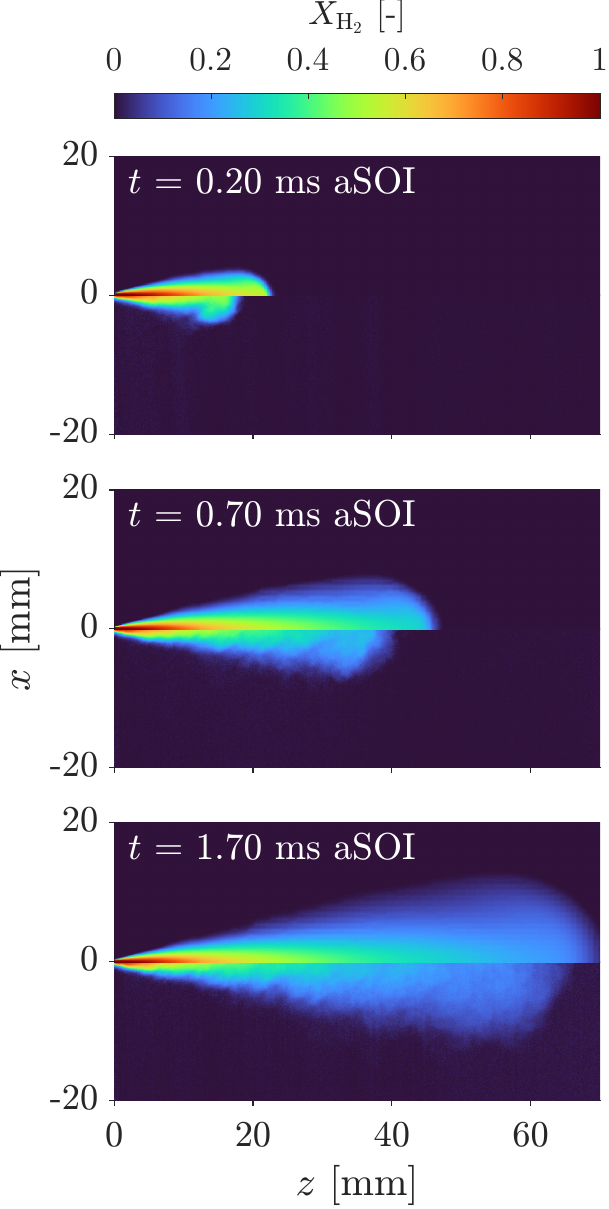}
      \caption{Standard $k$-$\varepsilon$ model.}
    \end{subfigure}
    \hfill
    \begin{subfigure}[t]{0.32\textwidth}
      \includegraphics[width=\textwidth]{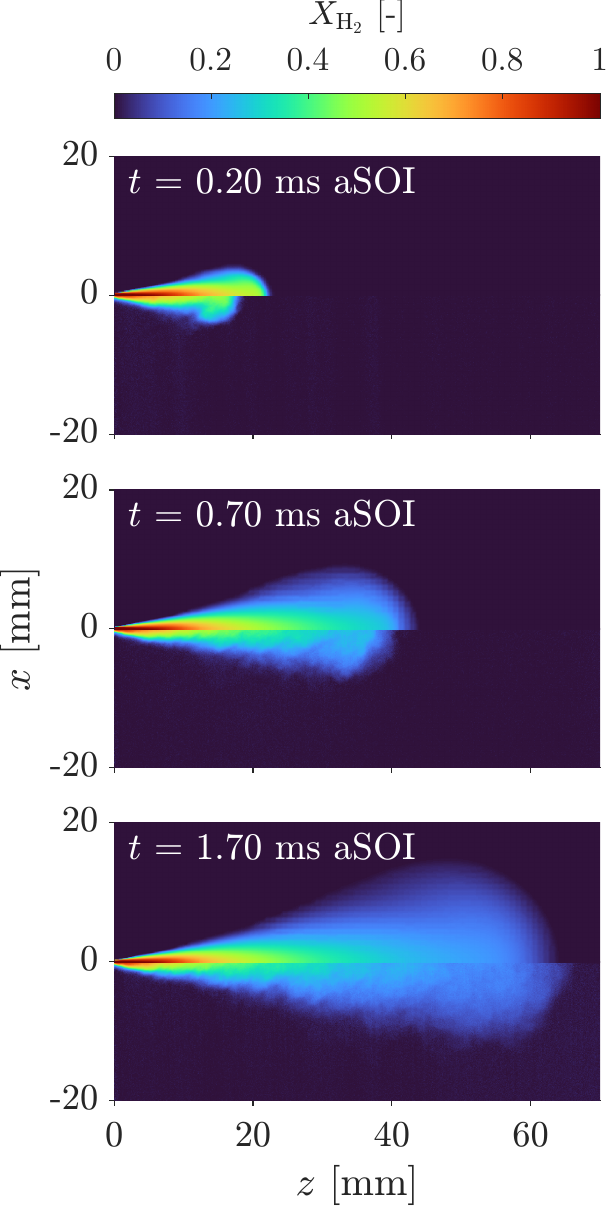}
      \caption{$k$-$\varepsilon$ RNG model.}
    \end{subfigure}
    \hfill
    \begin{subfigure}[t]{0.32\textwidth}
      \includegraphics[width=\textwidth]{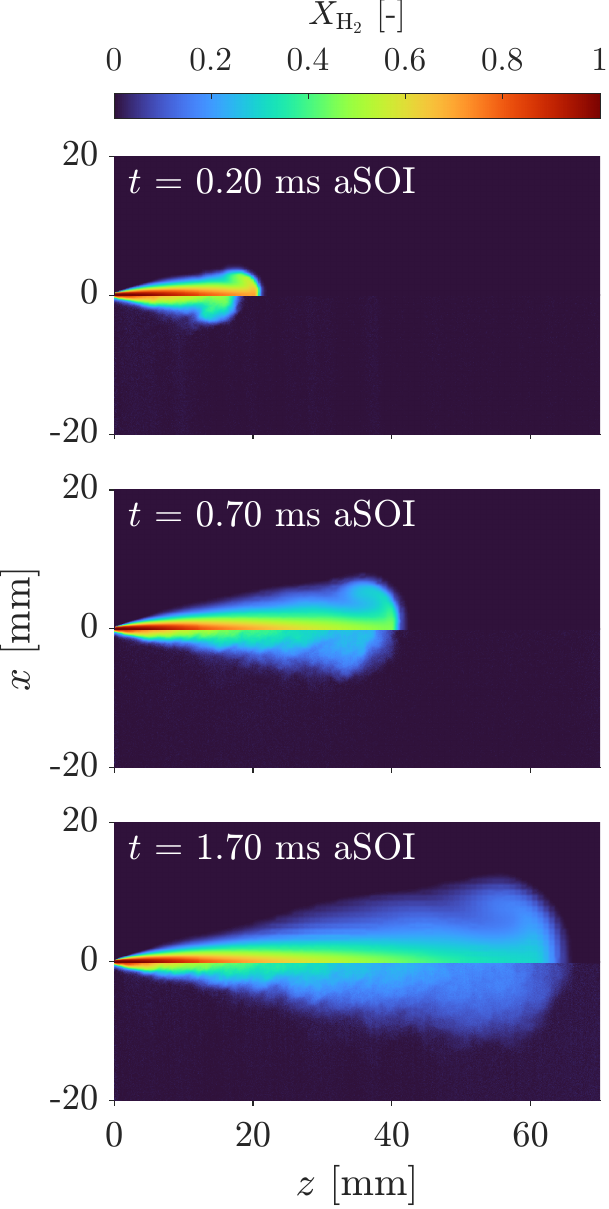}
      \caption{Reynolds stress model.}
    \end{subfigure}
    \caption{Comparison of distributions of mole fraction of hydrogen between the simulations (upper half jets) and the ensemble averaged ILIF measurements (lower half jets) at various times.}
    \label{fig:contours}
\end{figure*}

Figure~\ref{fig:contours} presents distributions of \ce{H2} mol fractions of the simulation results and ensemble averaged ILIF measurements at three different time instances after start of injection (aSOI).
The experimental images shown are from the incident laser side.
All three simulation results show reasonable agreement in regard to the temporal evolution and spatial distribution of the hydrogen jet, but differences are observed between the models.
The standard \keps model captures the shape of the jet well, but over-estimates the penetration depth of the jet tip consistently.
The \keps RNG model also over-estimates the jet tip penetration at \SI{0.20}{\milli\s}, but the over-estimation reduces at \SI{0.70}{\milli\s} and eventually leads to a minor under-estimation at \SI{1.70}{\milli\s}.
The angle of the \keps RNG jet at \SI{1.20}{} and \SI{1.70}{\milli\s} is wider compared to the experiment and standard \keps result.
This indicates that more mass is dispersed radially, which explains the slower penetration during this period.
Similar to the \keps models, the RSM over-estimates the jet penetration at \SI{0.20}{\milli\s}.
However, the agreement with the experiment improves over time. 
The shape of the jet is captured accurately as well by the RSM. 
The RSM is the only model that shows a vortex at the side of the jet's head. 
This vortex is not observed in the ensemble averaged ILIF results.

\begin{figure}
    \centering
    \includegraphics[width=0.7\textwidth]{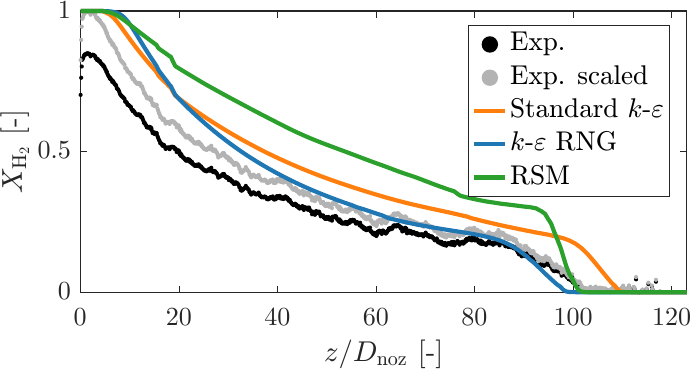}
    \caption{Axial profiles at centerline ($x=0$) of the experiment and the simulation results at \SI{1.7}{\milli\s} aSOI.}
    \label{fig:axial_profs}
\end{figure}

\begin{figure*}
  \begin{subfigure}[b]{0.32\textwidth}
    \includegraphics[width=\textwidth]{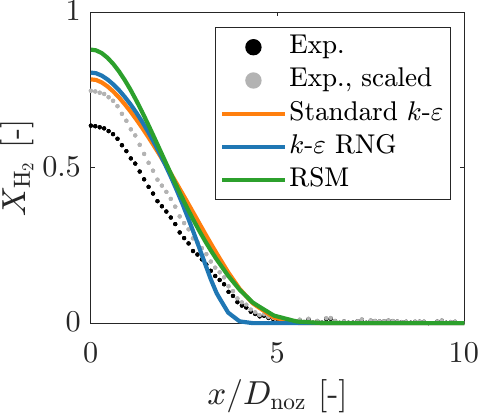}
    \caption{$z=\SI{10}{\milli\m}$}
  \end{subfigure}
  \hfill
  \begin{subfigure}[b]{0.32\textwidth}
    \includegraphics[width=\textwidth]{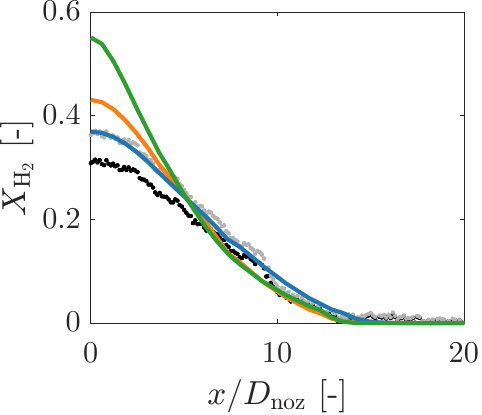}
    \caption{$z=\SI{30}{\milli\m}$}
  \end{subfigure}
  \hfill
  \begin{subfigure}[b]{0.32\textwidth}
    \includegraphics[width=\textwidth]{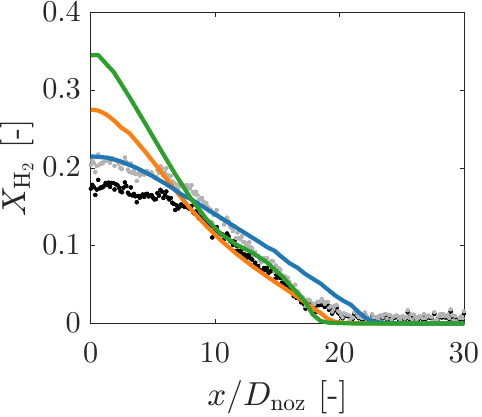}
    \caption{$z=\SI{50}{\milli\m}$}
  \end{subfigure}
  \caption{Radial profiles of the experiment and simulations at various axial distances at \SI{1.7}{\milli\s} aSOI.}
  \label{fig:radial_profs}
\end{figure*}

The experimental data shows lower mole fraction values almost everywhere in the jet.
This should not be the case, since the mass injection rates in the simulation and experiments are the same.
For this reason, the experimental hydrogen mole fractions are amplified to compare trends.
By assuming axi-symmetry of the half jet distributions of Fig.~\ref{fig:contours} and uniform molar density, the amount of moles of hydrogen in the vessel can be quantified by calculating the volume-integral. 
The calculated amount of moles of the ILIF results is a factor 0.73 lower compared to the simulation.
However, the maximum hydrogen mole fraction encountered in the ILIF contour plots of Fig.~\ref{fig:contours} is 0.85.
Multiplying the ILIF data by a factor of $1/0.73\approx1.37$ would result in values higher than 1 near the nozzle, which is unphysical.
For this reason, the experimental data in Figures~\ref{fig:axial_profs} and~\ref{fig:radial_profs} is scaled by a factor $1/0.85\approx1.18$ as a first-order measure to better compare trends between the simulation and experiment.

Figures~\ref{fig:axial_profs} and \ref{fig:radial_profs} show profiles of hydrogen mole fraction along the jet axis (i.e., where $x=0$) and along the radial direction at $z = [10,20,30]$ \unit{\milli\m}, respectively. 
Both scaled and unscaled experimental data are provided. 
The small steps in the simulated profiles of Fig.~\ref{fig:axial_profs} at $z/D_\mathrm{noz}\approx [15, 20, 75]$ are due to mesh refinements. 
The profiles along the jet axis presented in Fig.~\ref{fig:axial_profs} shows that the \keps RNG model results matches the scaled experimental data at the downstream part ($z/D_\mathrm{noz}>40$).
Hydrogen mole fractions upstream this part, on the other hand, are over-estimated.
The standard \keps model captures the trend of hydrogen decay, but consistently over-estimates the hydrogen mole fraction by approximately 0.07.
The RSM result does not compare well with the experimental data in terms of trend and absolute values, showing a difference with the experimental data of approximately 0.2.
Notably, all simulation results show a plateau of approximately \SI{5}{\milli\m} long where $X_{\ce{H2}}=1$, whereas the decay in the experimental data starts around \SI{3}{\milli\m}.
Inaccuracies near the nozzle exit are not surprising.
The nozzle exit flow is highly turbulent, but the ambient gas is not.
The flow in the vicinity of the nozzle can thus be characterized as a turbulent/non-turbulent interface.
A RANS simulation cannot capture the physics of such boundary layer, even with a second-moment closure model~\cite{gampert2014experimental}.

The profiles provided in Fig.~\ref{fig:radial_profs} show that the radial distribution of the hydrogen near the jet axis is best captured by the \keps RNG model. 
The tails of the profiles (where $X_{\ce{H2}}\lessapprox 0.1$), on the other hand, are better simulated by the other two models.
The trend of the profiles is captured reasonably well by the standard \keps model, especially at the tails. 
The RSM also captures the tails well, but shows large deviations near the jet axis.
Due to the low stoichiometric mixture fraction of hydrogen, ignition and combustion happens where hydrogen mole fractions are low.
Besides, the combustion process of a turbulent diffusion flame is limited by turbulent mixing, which mainly depends on the gradient of mixture fraction (by the scalar dissipation rate) rather than mixture fraction itself.
For these reasons, accurate simulation of the tails of the profiles may be more important than accurate prediction of the hydrogen distribution near the jet axis for simulation of auto-igniting jets.

\begin{figure*}
  \begin{subfigure}[b]{0.32\textwidth}
    \includegraphics[width=\textwidth]{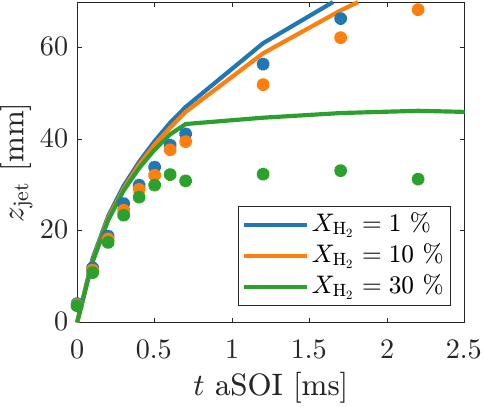}
    \caption{Standard $k$-$\varepsilon$ model.}
  \end{subfigure}
  \hfill
  \begin{subfigure}[b]{0.32\textwidth}
    \includegraphics[width=\textwidth]{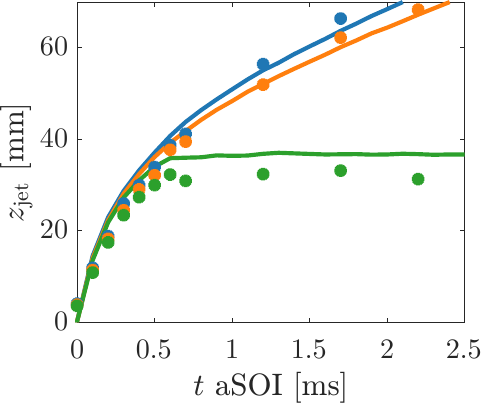}
    \caption{$k$-$\varepsilon$ RNG model.}
  \end{subfigure}
  \hfill
  \begin{subfigure}[b]{0.32\textwidth}
    \includegraphics[width=\textwidth]{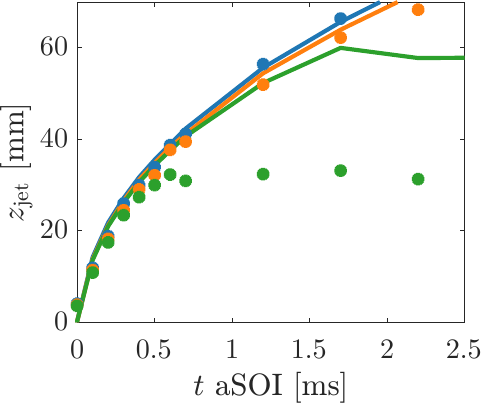}
    \caption{Reynolds stress model.}
  \end{subfigure}
  \caption{Jet penetration, denoted by $z_\mathrm{jet}$, of the scaled experimental data and the simulation results based on different levels of hydrogen mole fraction versus time.}
  \label{fig:penetration}
\end{figure*}

Another important characteristic of jet flames in ICEs is their penetration, because it is related to the flame surface area and affects flame-wall interaction, which is associated with heat loss in the engine.
Penetration curves based on different mole fraction criteria versus time are shown in Fig.~\ref{fig:penetration}. 
Among the three tested models, the \keps RNG model provides the best overall agreement, especially regarding the profile depicting $X_{\ce{H2}}=0.3$.
The standard \keps model consistently over-estimates the penetration for the three curves.
Still, its \SI{30}{\percent} curve is captured more accurately compared to the curve of the RSM.
The \SI{1}{\percent} curve of the RSM, on the other hand, compares best with the experiment among the turbulence models.
As mentioned earlier, all turbulence models over-estimate the jet penetration shortly after start of injection.
The differences can be due to a number of model simplifications. 
For example, complex flow phenomena during the needle opening transient are simplified by a linear process in the mass flow rate, pressure and temperature profiles.
Besides, the high Reynolds number limit assumption of the turbulence models may not be satisfied during the initial period.

\section{Reacting jets}
\label{sec:reacting_jets}
In this section, we assess the validity of the RANS setup to simulate auto-igniting hydrogen jets in \ce{Ar}/\ce{O2} and \ce{N2}/\ce{O2} environments for development of DI CI \ce{H2} engines.
For this purpose, simulation results are compared to pressure profiles as function of time which are measured by the closed vessel experiments of Peters et al.~\cite{peters2025thesis,peters2025ehpc}.
From these pressure profiles, combustion characteristics are derived, such as the ignition delay time and heat release.
In Section~\ref{sec:pressure_data_analysis}, this post-process routine is elaborated.
Temperature is not measured during these experiments but determined using the ideal gas law and is therefore not known exactly.
The volume-averaged temperature, also called bulk temperature, can vary among the different measurement cycles at a targeted condition due to variations occurring in the pre-burn event. 
Besides, temperature stratification occurs due to buoyancy and internal flow.
The so-called core temperature differs from the bulk temperature, which may be $\pm\SI{30}{\K}$.
Since the ignition delay time is sensitive to the local temperature, the effect of the uncertainty of temperature on the ignition delay is investigated in Section~\ref{sec:ignition_delay_time}.
Subsequently, simulation results obtained with the selected turbulence models at the baseline condition are compared to the corresponding experimental results for both ambients in Section~\ref{sec:baseline_condition}. 
Lastly, the capability of the CFD model to capture trends due to parametric variations in ambient temperature and oxygen concentration is investigated in Section~\ref{sec:parametric_variations}.

\subsection{Post-processing of pressure data}
\label{sec:pressure_data_analysis}
To validate the CFD model, it is important that the data from the experiments and simulation used for comparison include the same physical phenomena.
In the experiments, the pressure reduces due to heat loss.
This is not the case in the simulations since the boundaries of the vessel are modelled as adiabatic walls (see Section~\ref{sec:computational_domain}).
Except for heat loss, the CFD model should contain all effects that contribute to a change in pressure.
Therefore, to compare pressure rise profiles, the experimental data is corrected for heat loss.
This is realized by fitting first order polynomials to the raw pressure data before injection (between $t=[-10,0]$ \unit{\milli\s} aSOI), and after combustion ($t=[30,50]$ \unit{\milli\s} aSOI).
Although a power law fit is more accurate from a theoretical perspective, the difference in accuracy between linear and power law fits was found to be negligible, which can be explained by the moderate change in pressure during the considered interval.
The pressure loss rate before combustion is lower than after combustion, which is mainly due to an increase in temperature.
To create a continuous pressure loss profile during combustion that is related to the combustion process, the pressure loss rates before and after combustion are interpolated by using a combustion progress variable.
This variable is obtained by normalizing the pressure rise profile that is corrected for heat loss occurring after combustion.
By applying this method, it is assumed that the heat loss rate is proportional to the change in pressure.
Notably, the sampling rate of the simulation data is \SI{50}{\kilo\hertz}, which is equal to the experiments.

Besides pressure rise, heat release is also an important combustion characteristic that can be derived from the measured pressure profiles. 
To extract heat release by combustion, the first law of thermodynamics for open systems is used,
\begin{equation}
  \label{eq:1st_law}
  \di E = \delta Q - p \di V - \sum_{s=1}^{N_s}  \left(h_s \di m_s\right) \quad \text{ for } \quad s \in [1,N_s],
\end{equation}
where $\di E$ is the change in internal energy of the system, $\delta Q$ the heat transferred to the system, $p \di V$ the work done by the system and $\sum_{s=1}^{N_s} h_s\di m_s$ the enthalpy change due to mass exchange of species $s$ across the system boundary.
Considering that no work is done except for the injection of hydrogen and assuming ideal gas behavior, Eq.~\ref{eq:1st_law} can be rewritten to yield the heat release due to chemical reactions,
\begin{equation}
  \label{eq:hrr}
  \di Q_\mathrm{ch} = \left(e_{\ce{H2},\mathrm{ves}}-h_{\ce{H2},\mathrm{noz}}\right) \di m_\mathrm{inj}  + \sum_{s=1}^{N_s} \left(m_s \di e_{s,\mathrm{ves}}\right) - \frac{V}{\gamma - 1} \di p_\mathrm{HL} .
\end{equation}
where $e$ is the mass specific internal energy, $m_\mathrm{inj}$ the injected mass, $\di m_s$ the mass of species $s$ in the vessel, $\gamma$ the mixture-averaged specific heat ratio of the gases inside the vessel, and $\di p_\mathrm{HL}$ the change in pressure due to heat loss, which is determined by the method described above.
Subscripts ``noz'' and ``ves'' indicate nozzle and vessel conditions.
These properties are listed in Tables~\ref{tab:nozzle_conditions} and \ref{tab:ehpc_vessel_conditions}, respectively.
Similar to the heat loss profile, the internal energies and the masses of the different species, as well as the specific heat ratio are assumed to change in time with the combustion progress variable.
The time-dependent profile of $h_{\ce{H2},\mathrm{noz}}$ is created using the injection profiles as described in Section~\ref{sec:injection_conditions}.

\subsection{Ignition delay time}
\label{sec:ignition_delay_time}
First, the capability of the CFD model to predict ignition delay time (IDT) is investigated with the argon baseline condition.
Since the IDT is the time between SOI and start of combustion (SOC), it determines the amount of fuel available for premixed combustion.
Premixed combustion in compression ignition engines typically results in a high pressure rise rate, which can shorten the engine's lifetime or even lead to engine failure.
Increasing the in-cylinder temperature reduces the IDT hence the portion of premixed combustion, but it increases convective heat losses.
Therefore, to explore an engine's envelope, it is important to predict the IDT accurately.
In the experimental study of Peters et al.~\cite{peters2025thesis,peters2025ehpc}, it was concluded that SOC can be robustly determined based on a pressure rise of \SI{4}{\kilo\pascal}.
For this reason, the same definition is used here.

\begin{figure}
  \centering
  \begin{subfigure}[t]{0.6\textwidth}
    \includegraphics[width=\textwidth]{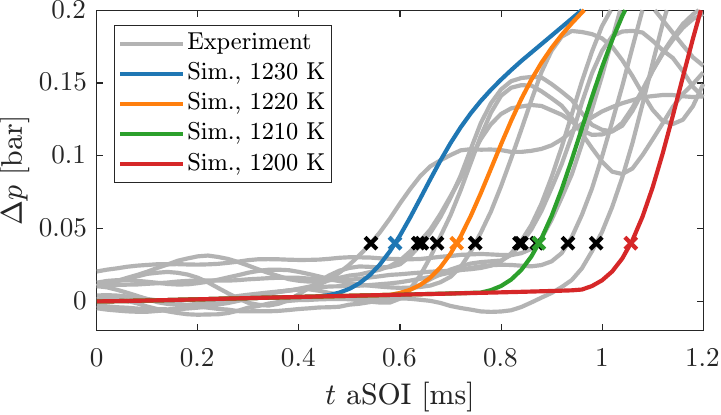}
    \caption{Pressure rise profiles. The simulated profiles are obtained with the RSM. The markers indicate start of combustion, which is determined based on a pressure rise of \SI{0.04}{\bar}.\label{fig:dp_plot}}
  \end{subfigure}
  \hfill
  \begin{subfigure}[t]{0.6\textwidth}
    \includegraphics[width=\textwidth]{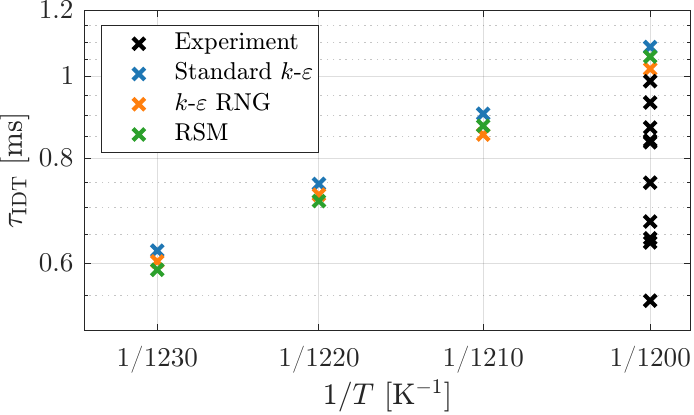}
    \caption{Arrhenius plot of ignition delay times. The experimental datapoints are plotted at the nominal temperature, but the uncertainty is \SI{30}{\K}.\label{fig:id_plot}}
  \end{subfigure}
  \caption{Pressure rise profiles versus time and ignition delay time versus initial vessel temperature of the simulations and experiments conducted at the argon baseline condition ($T=\SI{1200}{\K}$, $X_{\ce{O2}}=0.15$).}
\end{figure}

Figure~\ref{fig:dp_plot} presents pressure rise profiles of the experiments corrected for heat loss, and of the simulations. 
Both the experimental and simulation results are sampled at the same frequency (\SI{50}{\kilo\hertz}) and filtered using a Gaussian filter with a window of 25 datapoints. 
This filter was compared to various other methods, viz.\@ Savitzky-Golay, moving average, LOESS, LOWESS, Butterworth.
Based on this comparison, it was concluded that the Gaussian filter effectively reduced the noise, allowing for a smaller filter window than the other methods.
For readability of the plots, only the results obtained with the RSM are presented.
The simulations results shown in Fig.~\ref{fig:id_plot} are conducted at varying temperatures and demonstrate that the ignition delay time reduces with increasing temperature.
It is observed that SOC occurs in all measurements between 0.5 and \SI{1}{\milli\s}.
The CFD model is able to predict the IDT within this window, provided that the temperature in the computational domain is increased by at least \SI{10}{\K}.

Figure \ref{fig:id_plot} shows ignition delay times versus temperature of the experiments and simulations conducted at varying vessel temperatures.
Note that the experimental IDTs are plotted at the nominal temperature of \SI{1200}{\K}, but the range of uncertainty is \SI{30}{\K}.
The simulated IDTs show a clear proportionality in the Arrhenius plot.
Besides, the sensitivity of the turbulence models on the IDT is low.
The simulations performed at \SI{1200}{\K} over-predict the IDT.
The deviation in IDT simulated with the \keps RNG model and the highest experimental value is \SI{0.03}{\milli\s} or \SI{3.5}{\percent}.
The difference with the median of the experiments is obviously larger: \SI{0.23}{\milli\s} or \SI{29}{\percent}.
If the temperature in the computational domain is elevated by 10 or \SI{20}{\K}, which is well within the earlier explained uncertainty range, good agreement is found with the experimental values.

It is good to note that the inaccuracy in IDT does not necessarily have to be caused by the turbulence model or the uncertainty in nominal temperature, but also by the chemical mechanism.
The mechanism used in this study was validated on IDT using shock tube experiments, and the simulated IDTs around a temperature of \SI{1200}{\K} were typically \SI{5}{} to \SI{10}{\micro\s} longer than the experimental counterparts~\cite{burke2012comprehensive}.
These differences are not large enough to explain the deviations observed in Fig.~\ref{fig:id_plot} by the chemical mechanism alone.

There is also an uncertainty in composition in the experiments, mostly in the combustion products stemming from the pre-burn.
To gain feeling for the sensitivity of these products to the ignition delay, a simulation was run in which the \ce{CO2} was replaced by argon.
The difference in ignition delay was less than \SI{1}{\percent}.
Therefore, it is expected that the variations in oxidizer composition do not affect the combustion process considerably.
All other simulations are conducted with the compositions noted in Table~\ref{tab:ehpc_vessel_conditions}.

\subsection{Effect of uncertainty in ambient temperature}
In previous section, it is found that the experimental ignition delay of the argon baseline condition is captured well by the CFD model if the temperature in the computational domain is elevated from \SI{1200}{\K} to \SI{1210}{} - \SI{1230}{\K}.
In this section, we will discuss how the temperature affects heat release rate profiles based on the argon baseline condition and employing the RSM in the simulations.

\begin{figure}
  \centering
  \includegraphics[width=0.7\textwidth]{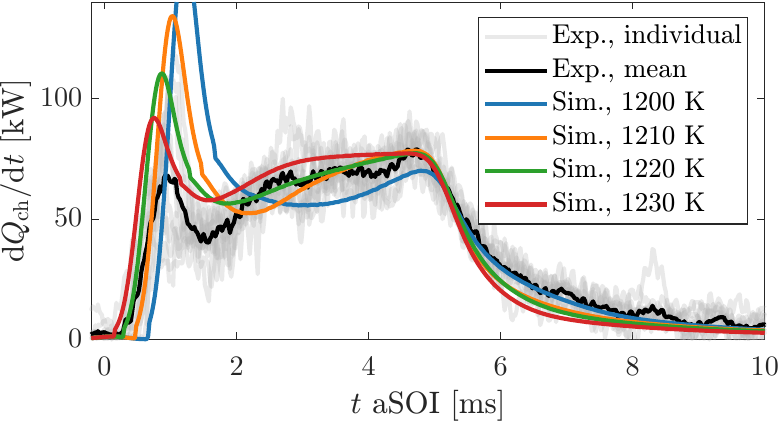}
  \caption{Heat release rate versus time of the experiments (Exp.) at the argon baseline condition ($T=\SI{1200}{\K}$, $X_{\ce{O2}}=0.15$) and of the simulations (Sim.) conducted with the RSM at varying temperatures.}
  \label{fig:hrr_temps}
\end{figure}

Figure~\ref{fig:hrr_temps} presents heat release profiles of the experiments and simulations at varying temperatures.
All profiles are treated with a Gaussian filter using a window length of 50, corresponding to \SI{1}{\milli\s}.
It is good to note that the heat release requires a larger filter window than the pressure rise profiles to suppress most of the noise due to the time-differentation.
The spikes between 0 and \SI{1.5}{\milli\s} seems to be what we know from diesel combustion as the premixed peak.
The period after \SI{1.5}{\milli\s} can be characterized as the diffusive, or mixing-controlled combustion phase.
As expected, the shorter IDT due to a small increase in ambient temperature results in lower premixed peak.
A similar trend was observed for the other two turbulence models.
The case simulated at \SI{1220}{\K} provides best agreement during the diffusive combustion phase following the premixed peak.
The height of the premixed peak is overestimated compared to the mean of the experiments. 
However, although difficult to see in the graph, some premixed peaks of the individual measurements exceed the simulated premixed peak.
Based on these observations and considering the uncertainty of the experiments, it is chosen to elevate the nominal temperature in the computational domains by \SI{20}{\K} from this point onward. 

\subsection{Baseline condition}
\label{sec:baseline_condition}
\subsubsection{Pressure rise}
\begin{figure}
  \centering
  \includegraphics[width=0.6\textwidth]{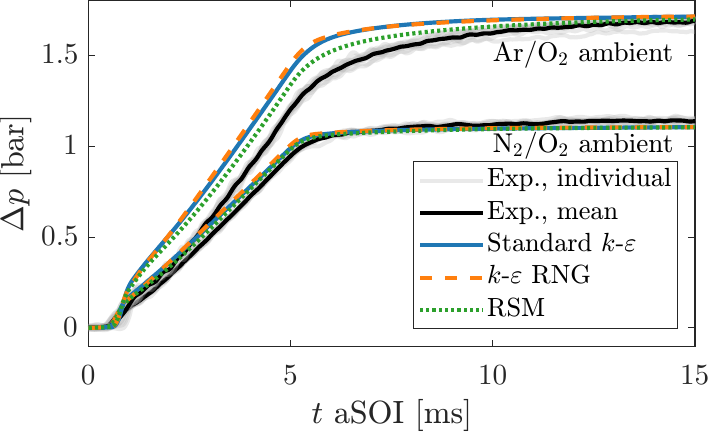}
  \caption{Pressure rise versus time of the experiments (Exp.) and simulations with the different turbulence models at the baseline condition ($T=\SI{1200}{\K}$, $X_{\ce{O2}}=0.15$).}
  \label{fig:dP_turb_models}
\end{figure}

To assess the performance of the different turbulence models, pressure rise profiles plotted against time in the \ce{Ar}/\ce{O2} and \ce{N2}/\ce{O2} ambients are presented in Fig.~\ref{fig:dP_turb_models}.
It is observed that the simulations overestimate the pressure rise during the injection, but converge to the experimental value towards the end of combustion.
This overestimation mainly happens around \SI{1}{\milli\s} aSOI, which is during the premixed combustion phase. 
After that, the profiles are nearly parallel.
Among the different turbulence models, the RSM provides the best agreement which is mainly because of the slower pressure rise near \SI{5}{\milli\s} in the \ce{Ar}/\ce{O2} ambient.
An analysis of the combustion process will be provided later this section.
The pressure rise in the argon case is clearly larger, despite the fact that the injection properties are the same for both ambients.
This is can be explained by argon's lower heat capacity ($c_V$) compared to nitrogen, which leads to a higher temperature and thus pressure rise for a certain heat release.
The CFD model captures this behavior well: the difference with the experimental mean pressure rise after \SI{15}{\milli\s} is \SI{3}{\percent} for the \ce{N2}/\ce{O2} ambient and less than \SI{1}{\percent} for the \ce{Ar}/\ce{O2} case.
Since $\Delta p \propto (\gamma -1)$ and given that the specific heat ratios of the \ce{Ar}/\ce{O2} and \ce{N2}/\ce{O2} ambients are 1.49 and 1.30, respectively, the pressure rise in nitrogen should be \SI{63}{\percent} of that in argon. 
This value is matched well by the simulations, but the experiments show a slightly larger difference of \SI{67}{\percent}. 
This deviation may be attributed to an uncertainty in the oxidizer composition having a direct effect on the specific heat ratio.

\subsubsection{Heat release}
\begin{figure}
  \centering
  \begin{subfigure}{0.6\textwidth}
    \includegraphics[width=\textwidth]{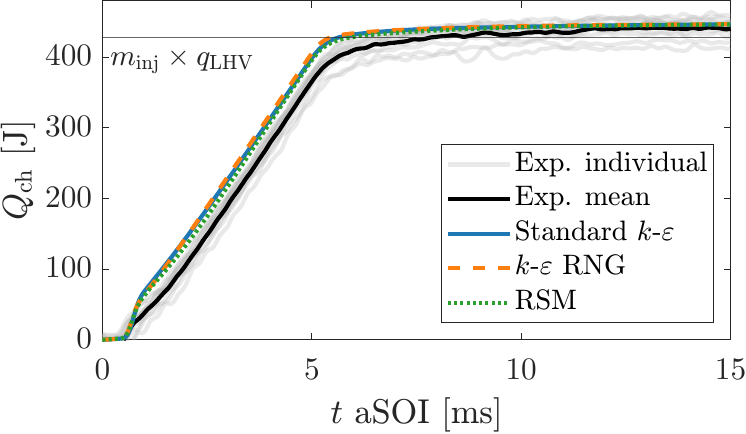}
    \caption{\ce{N2}/\ce{O2} ambient.}
    \label{fig:ihr_n2}
  \end{subfigure}
  \hfill
  \begin{subfigure}{0.6\textwidth}
    \includegraphics[width=\textwidth]{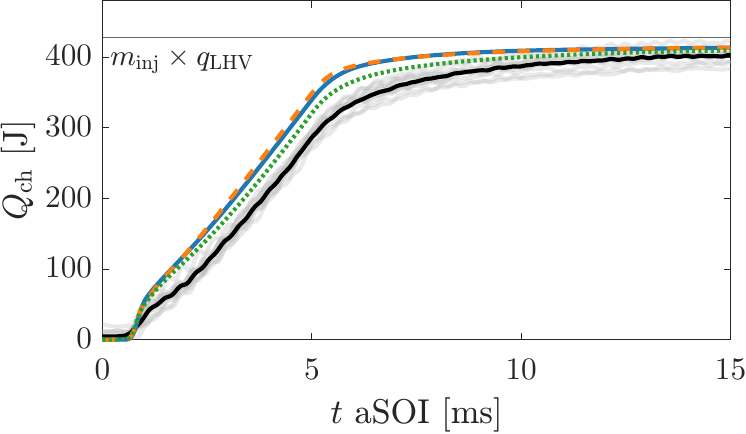}
    \caption{\ce{Ar}/\ce{O2} ambient.}
    \label{fig:ihr_ar}
  \end{subfigure}
  \caption{Heat release of the baseline condition versus time.}
  \label{fig:ihr_turb_models}
\end{figure}

Related to the pressure rise is the heat release, i.e.\@ the cumulative integral of the pressure-derived heat release rate, which is shown for the different turbulence models in Fig.~\ref{fig:ihr_turb_models}.
The theoretical total heat released, which is the product of the injected mass and the lower heating value of hydrogen based on the change in internal energy, is \SI{0.43}{\kilo\joule}.
This value is slightly over-predicted by the nitrogen cases and under-predicted by the argon cases.
The deviations in pressure-derived heat release rate may be explained by the different thermal equilibria that are reached if relatively cold hydrogen is mixed inertly with an \ce{N2}/\ce{O2} or \ce{Ar}/\ce{O2} ambient.
Because of the initial non-homogeneous temperature, the temperature and pressure at thermal equilibrium depend on the ambient composition.
In fact, for the inert case the pressure at thermal equilibrium for a nitrogen ambient is higher than the initial pressure, while in case of an argon ambient it will be lower.
This difference in pressure at thermal equilibrium is not captured by the pressure-derived heat release, since the method to derive it assumes that the gases inside the control volume are in thermal equilibrium.
This observation is consistent with the over- and under-prediction of the pressure-derived total heat release in Fig.~\ref{fig:ihr_turb_models} for the nitrogen and argon ambients, respectively.
A detailed investigation of this difference is left for further research.

\begin{figure}
  \centering
  \begin{subfigure}[b]{0.6\textwidth}
    \includegraphics[width=\textwidth]{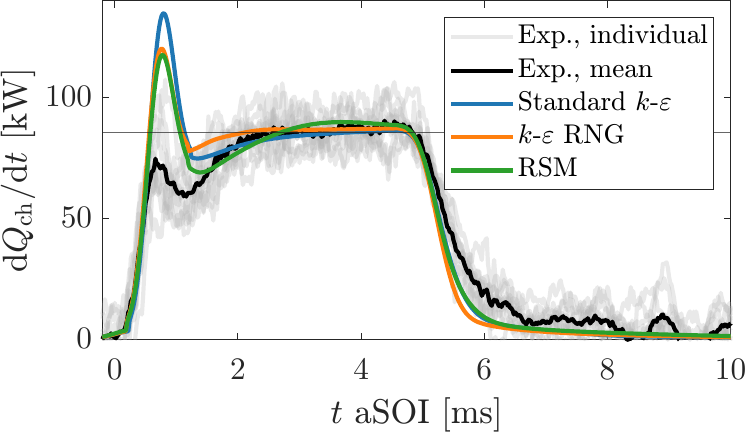}
    \caption{\ce{N2}/\ce{O2} ambient.}
    \label{fig:hrr_n2}
  \end{subfigure}
  \hfill
  \begin{subfigure}[b]{0.6\textwidth}
    \includegraphics[width=\textwidth]{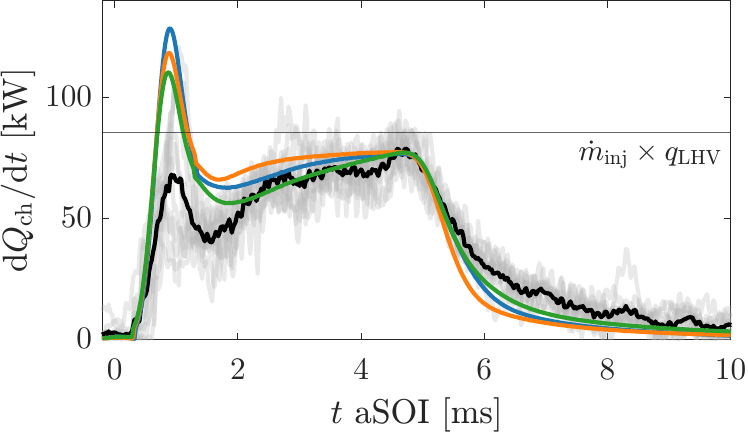}
    \caption{\ce{Ar}/\ce{O2} ambient.}
    \label{fig:hrr_ar}
  \end{subfigure}
  \caption{Pressure-derived heat release rate of the baseline condition versus time. The horizontal line indicates the energy injection rate during injection.}
  \label{fig:hrr_turb_models}
\end{figure}

\begin{figure}
  \centering
  \includegraphics[width=0.6\textwidth]{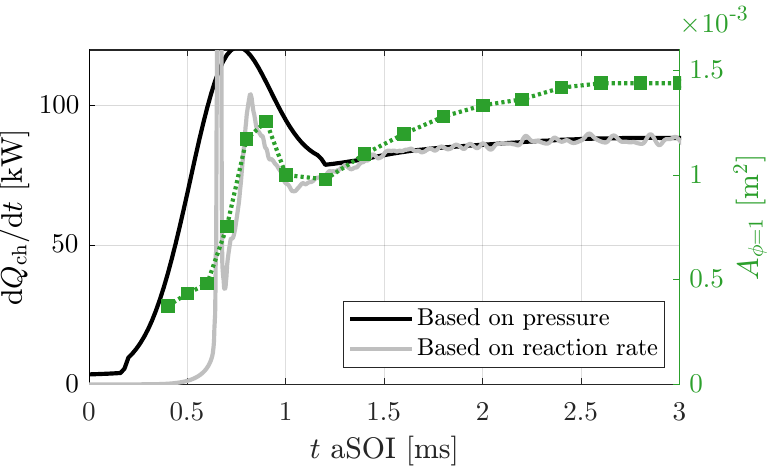}
  \caption{Heat release rates versus time of the simulation performed with the \keps RNG at the nitrogen baseline condition ($T=\SI{1200}{\K}$, $X_{\ce{O2}}=0.15$). The right $y$-axis depicts the surface area of the iso-surface of equivalence ratio $\phi=1$.}
  \label{fig:hrr_raw}
\end{figure}

\begin{figure}
  \centering
  \includegraphics[width=0.75\textwidth]{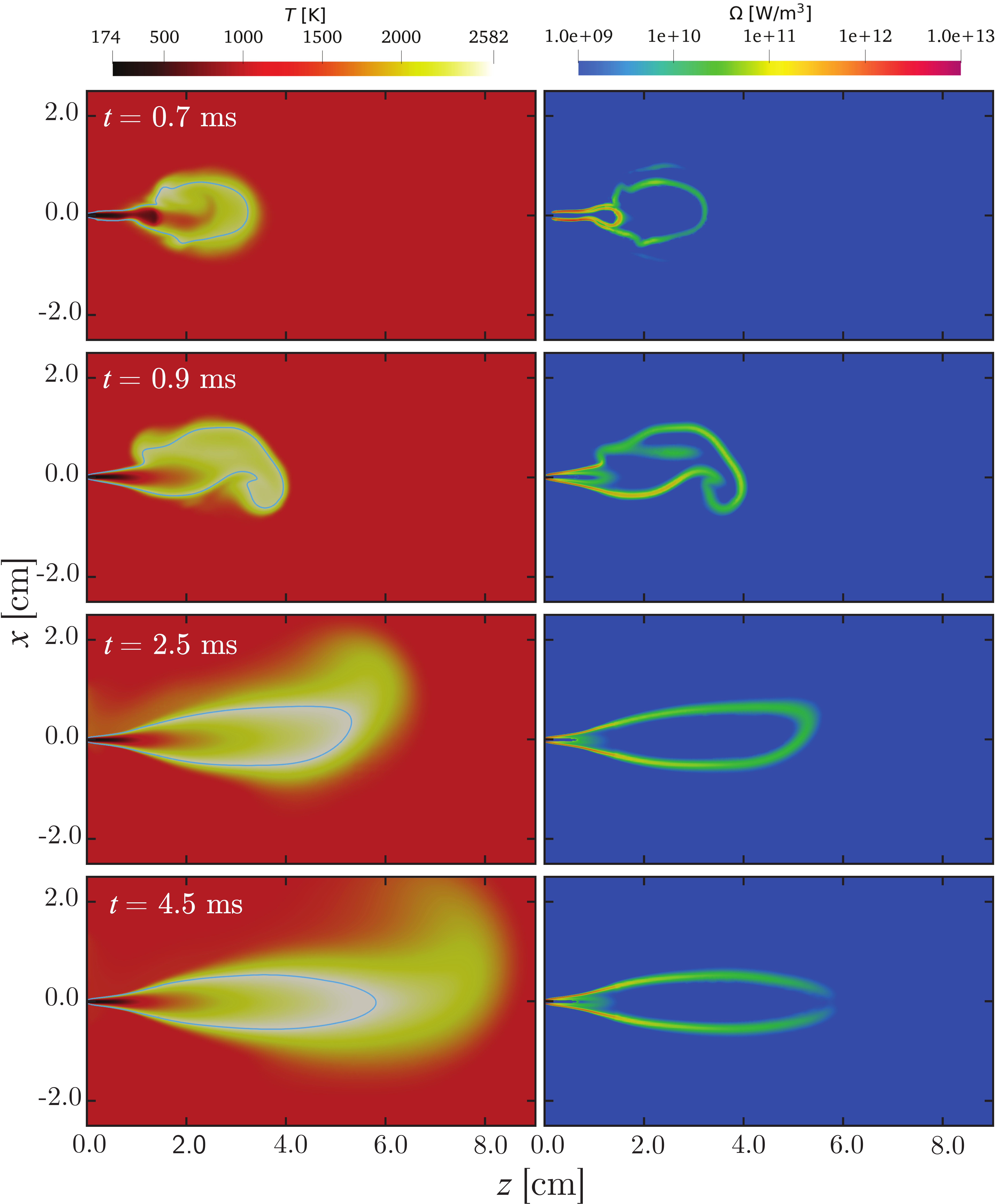}
  \caption{Contour plots showing temperature (left column) and the total chemical source term, $\Omega$, on logarithmic scale (right column) at various time instances. The blue iso-contours in the temperature plots depict equivalence ratio of 1.}
  \label{fig:chem_src_ignition}
\end{figure}

The differences between the experimental and simulated heat releases are explained in more detail based on the heat release rates shown in Fig.~\ref{fig:hrr_turb_models}.
The increase in heat release rate during the first \SI{0.5}{\milli\s} is captured well in the nitrogen case and reasonably well in the argon case. 
The height of the subsequent peak is overestimated when comparing to the mean of the experiments.
This is partly due to (small) differences in IDT between the individual experiments, leading to a lower and broader peak in the mean profile.
The average of the peak values of the individual experiments (gray profiles) is larger, \SI{85}{} and \SI{84}{\kilo\watt} for the nitrogen and argon case, respectively, but this does not explain the difference with the simulations.
To investigate the differences during this initial combustion phase, the pressure-derived heat release rate is compared to the one based on the reaction rate for the nitrogen case simulated with the \keps RNG model in Fig.~\ref{fig:hrr_raw}.
The reaction rate based heat release rate shows a narrow and high spike near \SI{0.7}{\milli\s}, followed by a second peak which reaches its maximum value at around \SI{0.8}{\milli\s}.
In the pressure-derived heat release rate, this second peak is retained while the narrow spike is filtered out.
Figure~\ref{fig:chem_src_ignition} presents contour plots showing the total chemical source term and temperature for the three turbulence models at various time instances during this initial combustion phase.
During the spike of the reaction rate based heat release rate in Fig.~\ref{fig:hrr_raw} between \SI{0.6}{} and \SI{0.7}{\milli\s}, the simulated jet ignites and a flame front engulfs the jet.
Since this ignition process happens so rapidly, it causes a perturbation in the flow field. 
This leads to an increased flame surface area and heat release rate at \SI{0.9}{\milli\s}, which is evident from the reaction rate based heat release rate in Fig.~\ref{fig:hrr_raw} and the contour plots showing the total chemical source term.
Note that although the computational problem is symmetric, the result after ignition is not.
An explanation for this could be that the computational grid is not symmetrically aligned to the nozzle exit plane.
In conjunction with the unstable nature of chemical kinetics, this could eventually lead to perturbations in the mean flow.
In the period between \SI{0.9}{} and \SI{1.0}{\milli\s}, the reactive zones at the side of the jet burn out, causing a temporary decrease in heat release, and the jet developes into a steady burning diffusion flame.
The surface area of the iso-surface of equivalence ratio of 1 is plotted versus time in Fig.~\ref{fig:hrr_raw}.
This profile shows that there is a clear correlation between flame surface area and reaction rate based heat release rate during this period.
It indicates that the peak in the pressure-derived heat release rate of the simulation is in fact mainly governed by diffusive combustion and that it is caused by the wrinkling of the flame surface area resulting from the ignition process.
A similar behavior is predicted by the other turbulence models.

It should be noted that in the experimental study of Yip et al.~\cite{yip2020visualization,yip2022parametric}, which concerns auto-igniting hydrogen jets in nitrogen-oxygen environments at similar conditions, a different mechanism is described.
They observed that the peak in heat release rate is governed by an ignition kernel that grows and engulfs the whole jet.
A similar mechanism is described in the numerical study of Ballatore et al.\cite{ballatore2025}, in which LES's of auto-igniting hydrogen jets are performed at the same conditions of \cite{yip2020visualization}.
The mechanism observed in the experiments and LESs corresponds well to what we know as the premixed peak from diesel spray combustion.
In the simulations presented here, this engulfment happens much faster during the initial spike of reaction rate based heat release rate.
This premixed peak is much sharper because turbulent fluctuations in temperature and composition are not accounted for. 
The heat release leads to a distortion of the flame front, which is not observed in the LESs nor in experiments.
Thus, although the peaks in pressure-derived heat release rate of the experiments and simulations around \SI{0.8}{\milli\s} appear relatively similar in Fig.~\ref{fig:hrr_turb_models}, they are governed by different phenomena.

After this initial combustion phase, the simulated heat release rates converge to the experimental ones during the diffusive combustion phase (see Fig.~\ref{fig:hrr_turb_models}).
The RSM provides accurate agreement with the mean experimental profile in both ambients, but also the two turbulent-viscosity models predict the trend rather well.
The heat release rate profiles in the nitrogen environment of the experiments as well as the simulations approach the energy injection rate ($\dot{m}_\mathrm{inj}\times q_\mathrm{LHV}$), meaning that the rates at which fuel is injected and at which it is converted to water are approximately equal.
This is especially obvious in the \keps RNG model result between \SI{2.5}{} and \SI{4.5}{\milli\s}.
This stagnation in heat release rate close to the injection rate implies that the jet has reached a quasi-steady state.
Figure~\ref{fig:chem_src_ignition} shows that this is indeed the case for the total chemical source term distributions, but not for the temperature distributions.
It indicates that the reactive zones reach a steady state and do not penetrate further than approximately \SI{60}{\milli\m}, while hot combustion products are still being transported downstream resulting in an increased flame penetration.
This implies that if the metal walls of the engine are at least \SI{60}{\milli\m} away from the nozzle, flame-wall interaction will result in heat loss but will not quench the chemical reactions.
The stagnation in heat release rate is not observed in the argon case.
Instead, it gradually increases until the injection stops.
These different trends are well-captured by the RSM and the standard \keps model.
The \keps RNG model slightly overestimates the heat release rate during the diffusive combustion phase of the argon case.
After the end of injection at \SI{5}{\milli\s}, the heat release rates decrease.
The experimental profile declines faster in the nitrogen case than in the argon case.
This difference is also captured by the CFD model.
In the simulations, however, the heat release rate ends more abruptly, which is most pronounced for the \keps RNG model.
The observations concerning the \keps RNG model can be related to the faster mixing behavior observed in Section~\ref{sec:non_reacting_jets}.
Interestingly, the difference in computational costs between the RSM cases and the ones simulated with the turbulent-viscosity models are negligible.

\subsection{Parametric variations}
\label{sec:parametric_variations}
In this section, we assess the capability of the CFD model to capture trends in heat release rate profiles due to variations in ambient temperature and oxygen concentration.
The simulations presented in this section are performed employing the RSM.
Furthermore, the ambient temperature is elevated by \SI{20}{\K} as earlier explained, which will affect the IDT less at a higher ambient temperature.

\begin{figure}
  \centering
  \begin{subfigure}[b]{0.6\textwidth}
    \includegraphics[width=\textwidth]{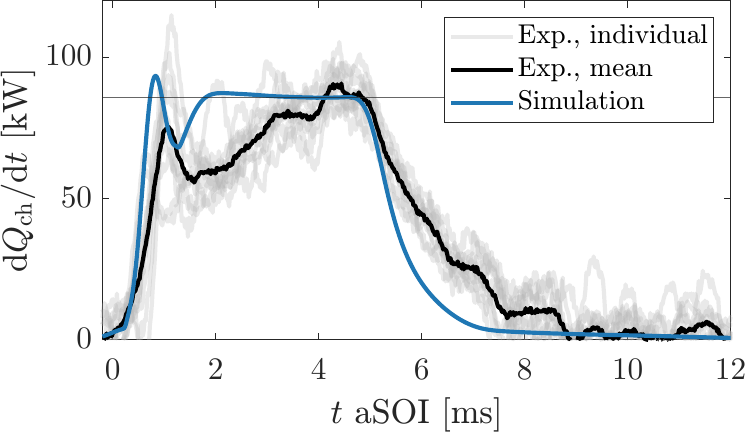}
    \caption{\ce{N2}/\ce{O2} ambient.}
    \label{fig:hrr_lowXn2}
  \end{subfigure}
  \hfill
  \begin{subfigure}[b]{0.6\textwidth}
    \includegraphics[width=\textwidth]{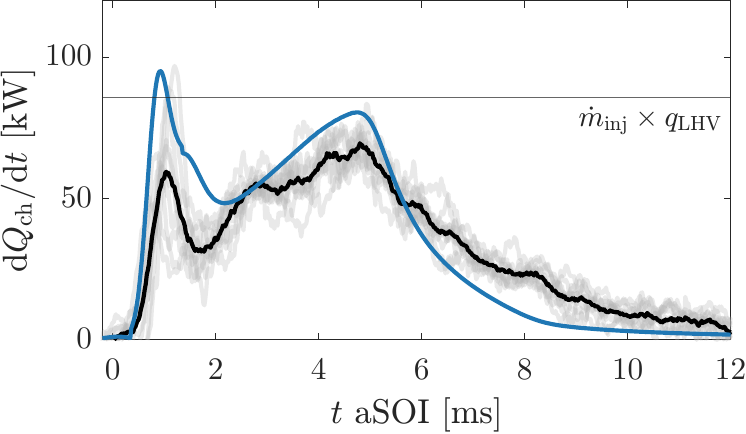}
    \caption{\ce{Ar}/\ce{O2} ambient.}
    \label{fig:hrr_lowXar}
  \end{subfigure}
  \caption{Heat release rate predicted by the RSM of the lower $X_{\ce{O2}}$ condition ($T=\SI{1200}{\K}$, $X_{\ce{O2}}=0.10$) plotted against time. The horizontal line indicates the energy injection rate during injection.}
  \label{fig:hrr_lowerXO2}
\end{figure}

The heat release profiles of the variation with a lower oxygen concentration of \SI{10}{\percent} are shown in Fig.~\ref{fig:hrr_lowerXO2}.
The start of combustion is well-predicted for both ambients, similar to the baseline condition.
Due to the lower oxygen concentration in the ambient, the oxygen entrainment rate is lower resulting in lower heat release rates compared to the previous conditions.
After end of injection, the heat release rates decrease less steeply, which can also be explained by the lower oxygen entrainment rate.
The CFD model captures these trends well for the argon condition, see Fig.~\ref{fig:hrr_lowXar}.
The differences with the experimental profiles are consistent with the ones of the baseline condition: the peak around \SI{1}{\milli\s} is sightly overestimated, accurate agreement is found during the diffusive combustion phase.
The simulated heat release rate profile for the nitrogen ambient, which is plotted in Fig.~\ref{fig:hrr_lowXn2}, shows a plateau similar to the baseline condition in the nitrogen ambient. 
This plateau is not observed in the experimental data.
On the other hand, the peak near \SI{1}{\milli\s} and the heat release rate towards the end of injection is captured accurately.

\begin{figure}
  \centering
  \begin{subfigure}[b]{0.6\textwidth}
    \includegraphics[width=\textwidth]{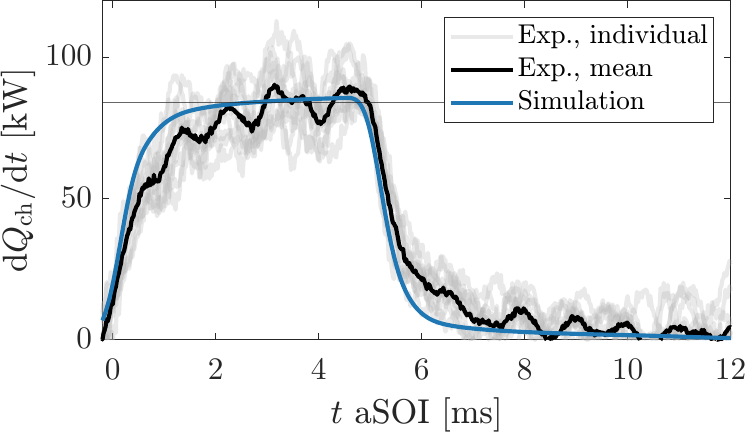}
    \caption{\ce{N2}/\ce{O2} ambient.}
    \label{fig:hrr_highTn2}
  \end{subfigure}
  \hfill
  \begin{subfigure}[b]{0.6\textwidth}
    \includegraphics[width=\textwidth]{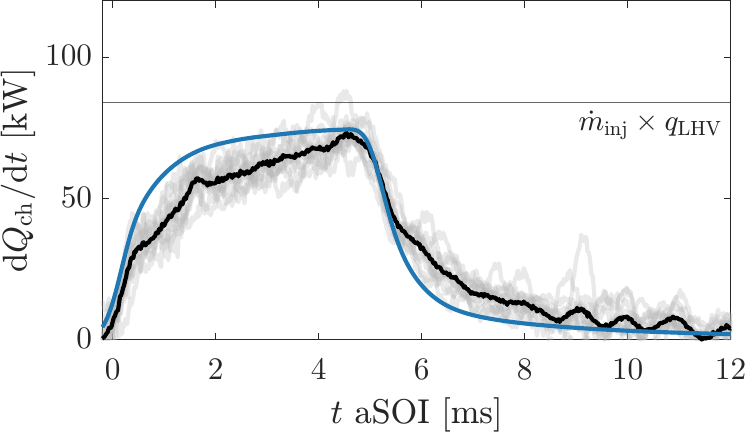}
    \caption{\ce{Ar}/\ce{O2} ambient.}
    \label{fig:hrr_highTar}
  \end{subfigure}
  \caption{Heat release rate predicted by the RSM of the higher $T$ condition ($T=\SI{1400}{\K}$, $X_{\ce{O2}}=0.15$) plotted against time. The horizontal line indicates the energy injection rate during injection.}
  \label{fig:hrr_highT}
\end{figure}

In Fig.~\ref{fig:hrr_highT}, heat release rate profiles of the higher $T$ condition ($T=\SI{1400}{\K}$, $X_{\ce{O2}}=0.15$) are presented.
Because of the high ambient temperature, the hydrogen jet ignites almost instantaneously resulting in a purely diffusive combustion process.
The experimental heat release rate of the argon case increases more slowly compared to the nitrogen case.
As a result, the heat release rate profile of the nitrogen ambient approaches the injection rate, while the one of the argon ambient clearly increases until the end of injection.
This is captured by the simulations to some extent, but in both ambients the simulated profiles approach a constant value resulting in a better agreement for the nitrogen case than for the argon case.
The heat release rate is overestimated by the CFD model in the argon case after \SI{0.5}{\milli\s}, but it has converged to the experimental profile at the end of injection.
During the burn-out phase after \SI{5}{\milli\s}, the heat release rates are underestimated by the simulations, which was also observed for the baseline condition.

\section{Conclusions}
\label{sec:conclusions}
In the present study, a RANS CFD model is setup in a widely used software package for simulation of high-pressure hydrogen injections at engine conditions.
Two turbulent-viscosity models (the standard \keps and the \keps RNG) and the Reynolds stress model (RSM) are compared against inverse laser-induced fluorescence experiments to validate the turbulent mixing behavior of non-reacting jets.
Overall, adequate agreement is found regarding the prediction of hydrogen distribution.
Trends in hydrogen jet distributions are adequately simulated using the standard \keps model.
The \keps RNG model provides good qualitative agreement in hydrogen distributions, especially in the head of the jet.
Jet tip penetration is most accurately predicted by the RSM.
This is an important characteristic for engine research, as the jet tip penetration determines flame-wall interaction and the associated heat losses.

The capability of the CFD model to simulate auto-igniting hydrogen jets is assessed by comparison of pressure traces obtained by closed-vessel experiments found in literature.
Accurate agreement is found in the pressure rise due to combustion.
This is an important characteristic for engine research as the pressure rise directly translates to work done by the engine.
The combustion model used in this dissertation does not account for turbulent fluctuations in temperature and composition, which is clearly a shortcoming from a theoretical perspective.
As a consequence, the premixed combustion phase happens too rapidly and is not predicted adequately. 
The simulated heat release rate profile which is derived from the pressure shows a peak similar to the premixed peak observed in the experiments, but that peak is governed by a different phenomenon.

Despite the neglection of turbulent fluctuations in temperature and composition in the combustion model, good agreement is found in heat release rate profiles during the diffusive combustion phase, especially with the RSM.
This indicates that, during this period, combustion is determined by turbulent mixing.
Adequate agreement is found for trends due to variations in ambient temperature and oxygen concentration.
To perform more detailed flame analyses, e.g.\@ regarding temperature distributions, a deeper model validation regarding local flame temperatures is required.

The presented RANS CFD model is an effective method to predict domain average variables, such as pressure and heat release rate, of auto-igniting high-pressure hydrogen injections and can be used to develop direct-injected compression-ignition hydrogen engines.
This study therefore aids the development of zero-emission internal combustion engines, such as the argon power cycle, and contributes to overcoming the challenge of the intermittent energy supply by renewable energy sources.

\section*{CrediT authorship contribution statement}
N. Diepstraten: Conceptualization, Formal analysis, Investigation, Methodology, Validation, Visualization, Writing -- original draft.
Bart Somers: Supervision, Writing -- review \& editing.
Jeroen van Oijen: Conceptualization, Funding acquisition, Project administration, Resources, Supervision, Writing -- review \& editing.

\section*{Declaration of competing interest}
The authors declare that they have no known competing financial
interests or personal relationships that could have appeared to
influence the work reported in this paper.

\section*{Acknowledgments}
This publication is part of the project Argon Power Cycle (with project number 17868) of the VICI research programme which is financed by the Dutch Research Council (NWO).

\bibliographystyle{unsrtnat}
\bibliography{references}

\end{document}